\documentclass[aps,prb,reprint,superscriptaddress,showpacs,showkeys]{revtex4-1}

\usepackage{graphicx}
\usepackage{dcolumn}
\usepackage{bm}
\usepackage{color}
\usepackage{soul} 

\newcommand{\eq}{\begin{equation}}
\newcommand{\eqn}{\end{equation}}
\newcommand{\vek}[1]{\bm{\mathrm{#1}}} 
\newcommand{\cmThree}{cm$^{-3}$}

\newcommand{\mum}{$\mu$m}

\newcommand{\algaas}{Al$_{x}$Ga$_{1-x}$As}

\begin{document}


\title[Sample title]{Fast optical control of spin in semiconductor interfacial structures}

\author{L.~N\'{a}dvorn\'{i}k}
 \email{nadvl@fzu.cz}
 \affiliation{Institute of Physics ASCR, v.v.i., Cukrovarnick\'{a} 10, 16253 Praha 6, Czech Republic}
 \affiliation{Department of Physical Chemistry, Fritz Haber Institute of the Max Planck Society, 14195 Berlin, Germany}
\author{M.~Sur\'ynek}
 \affiliation{Faculty of Mathematics and Physics, Charles University, Ke Karlovu 3, 12116 Praha 2, Czech Republic}
\author{K.~Olejn\'{i}k}
 \affiliation{Institute of Physics ASCR, v.v.i., Cukrovarnick\'{a} 10, 16253 Praha 6, Czech Republic}
\author{V.~Nov\'{a}k}
 \affiliation{Institute of Physics ASCR, v.v.i., Cukrovarnick\'{a} 10, 16253 Praha 6, Czech Republic}
 \author{J.~Wunderlich}
 \affiliation{Institute of Physics ASCR, v.v.i., Cukrovarnick\'{a} 10, 16253 Praha 6, Czech Republic}
 \affiliation{Hitachi Cambridge Laboratory, J. J. Thomson Avenue, CB3 0HE Cambridge, UK}
 \author{F.~Troj\'{a}nek}
 \affiliation{Faculty of Mathematics and Physics, Charles University, Ke Karlovu 3, 12116 Praha 2, Czech Republic}
\author{T.~Jungwirth}
 \affiliation{Institute of Physics ASCR, v.v.i., Cukrovarnick\'{a} 10, 16253 Praha 6, Czech Republic}
 \affiliation{School of Physics and Astronomy, University of Nottingham, Nottingham NG7 2RD, UK}
\author{P.~N\v{e}mec}
 \affiliation{Faculty of Mathematics and Physics, Charles University, Ke Karlovu 3, 12116 Praha 2, Czech Republic}



\date{\today}

\begin{abstract}
{We report on a picosecond-fast optical removal of spin polarization from a self-confined photo-carrier system at an undoped GaAs/AlGaAs interface possessing superior long-range and high-speed spin transport properties. We employed a modified resonant spin amplification technique with unequal intensities of subsequent pump pulses to experimentally distinguish the evolution of spin populations originating from different excitation laser pulses. We demonstrate that the density of spins, which is injected into the system by means of the optical orientation, can be controlled by reducing the electrostatic confinement of the system using an additional generation of photocarriers. It is also shown that the disturbed confinement recovers within hundreds of picoseconds after which spins can be again photo-injected into the system. 

} 
\end{abstract}

\maketitle



\section{Introduction}

Magnetic random access memory (MRAM) bits\cite{chappert2007} and spin Hall effect transistors\cite{wunderlich2010} are examples of two different conceptual approaches to spintronic devices whose functionality requires different spin-conserving length-scales. The operation of MRAM bits is based on a~vertical transfer of spins through a~nm-thin layer in a~``sandwich-like'' layered structure.\cite{wolf2001} In spin-logic devices, however, spins have to be manipulated (by electric gates, for instance) before they reach the drain electrode which calls for a~lateral design rather than the vertical geometry.\cite{datta1990,sarma2001,jalil2008} Orders of magnitude larger spin-conserving length-scale is, therefore, needed in the spin-logic devices due to the typical dimensions of the order of a \mum\ of lateral structures fabricated by the electron-beam lithography.\cite{wunderlich2009,wunderlich2010,nadvornik2015,nadvornik2016a} The desired high-rate operation of the devices implies an additional requirement on a fast spin transport over this length scale. The criteria for the simultaneously long-range and high-speed spin transport are not satisfied in a majority of standard systems, including $n$-doped bulk semiconductors\cite{kikkawa1998,dzhioev2002,sprinzl2010} or (001)-grown quantum wells\cite{wu2010}, due to their limited spin mobility or spin life time, respectively. Another critical parameter which limits the operation rate of a spin-logic device is the speed of spin manipulation. While the spin injection and detection can be a fast process (for instance by means of the optical orientation\cite{agranovich1984} and magneto-optical probing\cite{korn2010,wu2010}), a fast removal of the spin polarization from the system is a much more challenging task.


In this paper, we show experimentally that the electronic spin polarization can be removed from a transport channel of a self-confined system in a picosecond time-scale using a control optical pulse which reduces the level of its confinement. We demonstrate this novel functionality in an optically generated self-confined system formed at an undoped GaAs/\algaas\ heterointerface where the spin transport meets the other key requirements of high speed and long range, as reported recently in Ref.~\onlinecite{nadvornik2016}. In addition, we show that the confinement of the transport layer is recovered after the removal of the spin polarization within a few hundreds of picosecond. Combining all these favorable characteristics, we obtain a unique candidate system for the development of the spin logic concept in semiconductors.

Our study was performed by a modified resonant spin amplification (RSA) technique\cite{kikkawa1998} using unequal intensities of neighbouring pump pulses which enables us to probe simultaneously the time evolution of mixed spin populations created at different time instants. This scheme also allows us to generate the spin polarization by a lower intensity pulse in a well controlled way without disturbing the confinement, then to reduce it by a high intensity pulse, and eventually to read it by a probe pulse.

The paper is organized as follows: First, we describe the studied structure and explain the mechanism of the creation of the long-lived electronic spin sub-system and show its fingerprint in a~magneto-optical (MO) signal. Then we describe the modified RSA technique with unequal intensities of subsequent pump pulses. In the following sections, we discuss our experimental results demonstrating the fast optical control of spin polarization. Finally, in appendices, we provide additional information on our experimental setup and an estimate of the impact of the disturbed confinement on the spin polarization. 

\section{Studied system}
\subsection{Sample composition}\label{sec:sample}
\begin{figure}
\includegraphics[width=1.0\columnwidth]{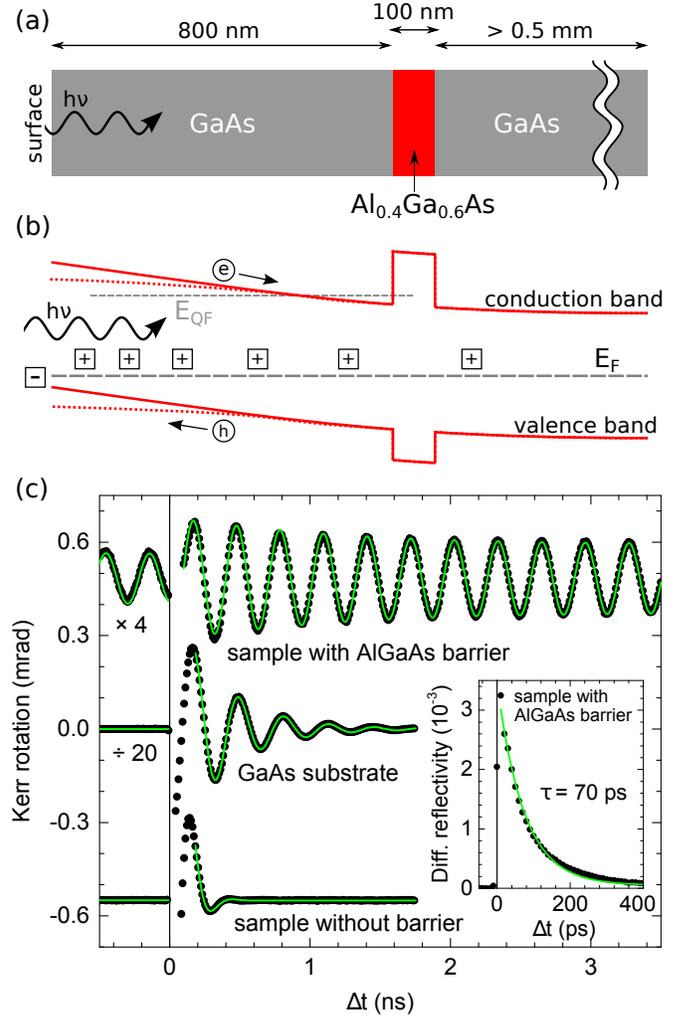}
\caption{\label{sample-Kerr} (Color online) (a)~The layer structure of the studied sample (the sample surface, on which the light is incident, is on the left side). (b)~The energetic diagram of the conduction and valence bands before (solid curves) and after the illumination (dotted curves). The Fermi level $E_F$ is pinned to the edge of the valence band due to the negatively charged surface states (the square with the minus sign) and ionized unintentional bulk impurity states (squares with the plus sign) and the band banding results in a built-in electric field. After the illumination by light with photon energy $h\nu =1.52$~eV, the photo-created electrons (circle with the letter \textit{e}) and holes (circle with \textit{h}) are driven by the built-in electric field and spatially separated (arrows). This process compensates partially the electric field and leads to a~formation of a~steady-state long-lived electron sub-system near the upper GaAs/AlGaAs interface, depicted by the quasi-Fermi level $E_{QF}$. The electron-hole overlap is minimal in the sub-system which suppresses the recombination and allows for long spin lifetime $\tau_s$. (c)~The measured dynamics of magneto-optical Kerr signal (points) with fits by Eq.~\ref{eq:RSA-all} (solid curves). The data measured in the corresponding samples at 10~K, $B=500$~mT, $h\nu=1.52$~eV, and 14~$\mu$J$\cdot$cm$^{-2}$ were multiplied or divided by the indicated factors and vertically shifted for clarity. The non-zero signal for $\Delta t<0$ in the sample with the AlGaAs barrier is a~signature that $\tau_s$ in the electron sub-system exceeds the time spacing between the neighbouring pump laser pulses $ t_0$, which is 12.5~ns in this particular case. Inset: The measured dynamics of the differential reflectivity (points) for the sample with the AlGaAs barrier. The depicted recombination time $\tau$ was inferred from an exponential fit (curve).}
\end{figure}

The studied long-lived and highly mobile electron spin-system is self-confined near the upper GaAs/\algaas\ interface in an undoped heterostructure.\cite{nadvornik2016} This spin system is formed in a~wide range of $x$-composition and layer thicknesses as a consequence of the spatial separation of optically generated electron-hole pairs in a built-in electric field due to surface states. In this paper, we use the optimal Al composition, $x=0.4$, and the most simple layer composition shown in Fig.~\ref{sample-Kerr}(a). A~100~nm thick undoped Al$_{0.4}$Ga$_{0.6}$As barrier was deposited by a molecular beam epitaxy on a top of an insulating GaAs buffer and a GaAs substrate. The barrier was then covered by another undoped 800~nm thick GaAs layer where the confined electron sub-system is formed, as explained below.

\subsection{Creation of electronic sub-system}\label{sec:separation}
The process which usually limits the spin lifetime $\tau_s$ in undoped structures is the electron-hole recombination.\cite{kikkawa1998} The origin of the unusually long $\tau_s$ measured in our undoped sample is the suppression of the recombination spin decay channel due to the presence of a long-lived sub-system of photo-generated excess electrons. This effective optically generated doping is possible due to the spatial separation of photo-generated electrons and holes in the built-in electric field. If we consider our structure in dark, charged residual bulk and surface states pin the Fermi level near the valence band\cite{nadvornik2016} which causes a~band bending and, thus, an electric field across the structure -- see Fig.~\ref{sample-Kerr}(b), solid curves. When the sample is illuminated from the surface side, photo-electrons and photo-holes are generated. If the structure did not contain the barrier, the photo-carriers would migrate in the built-in electric field, fill all the charged states and compensate fully the electric field. However, because of the AlGaAs barrier, the free migration between the layers above and below the barrier is obstructed and the full compensation of the field is not possible (dotted curves). The residual built-in electric field leads to a~formation of a~steady-state optically excited electronic system near the upper GaAs/AlGaAs interface, depicted by the quasi-Fermi level in Fig.~\ref{sample-Kerr}(b). Due to the minimal spatial overlap between electrons and holes, 
the carrier recombination is suppressed. The long-lived system created this way can be viewed as an effective (optically generated) local $n$-doping which suppresses the spin decay channel via the carrier recombination and allows for the observed long $\tau_s$, similarly as in $n$-doped GaAs samples\cite{dzhioev2002,kikkawa1998}. The process is described explicitly in Sec.~\ref{sec:discussion}; a~more detailed discussion with its experimental confirmation can be found in Ref.~\onlinecite{nadvornik2016} and its Supplementary information.

\subsection{Time-resolved magneto-optical detection}\label{sec:typical-kerr}
The presence of the steady-state long-lived electronic spin system in the studied sample is revealed by the magneto-optical (MO) pump-and-probe experiment; we use the optical orientation and the MO Kerr effect for spin injection and detection, respectively, as described in detail in Appendix~\ref{app:setup} and Fig.~\ref{setup}. In Fig.~\ref{sample-Kerr}(c) we show the data measured in the sample with the localized electronic sub-system and in two reference samples using laser pulses with a repetition frequency of 80~MHz, i.e., a time-spacing between neighbouring laser pulses $t_0=12.5$~ns. Clearly, the presence of the AlGaAs barrier is essential for the achievement of the long spin lifetime. In fact, even without any quantitative analysis, the presence of the MO signal at negative pump-probe time delays $\Delta t$ (i.e., the fact that the spin signal induced by a previous pump pulse did not decay to zero before the impact of an another pump pulse) immediately indicates that $\tau_s\gtrsim t_0=12.5$~ns. For comparison, the typical recombination time $\tau$ is of the order of hundreds of picosecond in the undoped GaAs.

The fitting of the data provides a more precise determination of $\tau_s$ as shown in Fig.~\ref{sample-Kerr}(c). 
By applying the model for the total MO signal presented in Ref.~\onlinecite{nadvornik2016} (i.e., Eq.~\ref{eq:RSA-all} for $m=0,1$ and $A_0=A_1=A$ below) on the data [solid curves in Fig.~\ref{sample-Kerr}(c)]
we deduced $\tau_s\approx 16$~ns
which is, indeed, more than two orders of magnitude larger than the carrier recombination time $\tau\approx 70$~ps, inferred from the dynamics of the differential reflectivity measured in this sample. From the dependence of the fitted oscillatory (Larmor) frequency on applied magnetic field we inferred the $g$-factor of magnitude $0.45\pm0.01$, confirming that the spin-carriers are free electrons.\cite{zawadzki2008} In the bare GaAs substrate and the undoped epitaxial GaAs on the substrate, the spin lifetimes are more than two orders of magnitudes shorter -- 350~ps and 70~ps for the substrate and epitaxial GaAs, respectively. This agrees well with the observed $\tau$ in the sample with the barrier and with typical values reported for undoped bulk GaAs where the spin lifetime of spin-polarized photo-carriers is limited by their recombination time and where $\tau$ is typically at the time-scale of 100's of ps are reported.\cite{kikkawa1998, wu2010,nemec2005}

We point out that the MO detection technique enables us to separate signals coming from different parts of the studied sample. In particular, a proper selection of the probe detection wavelength enhances the sensitivity to the detected MO signals originating from the localized electronic sub-system (see Appendix~\ref{app:setup}, Ref.~\onlinecite{nadvornik2016} and its Supplementary information for more details). Thanks to this, the measured MO dynamics is not dominated by contributions of short-lived spins generated in the bulk despite their much a higher density [see Fig.~\ref{sample-Kerr}(c)].

\section{RSA technique with unequal intensities of subsequent laser pulses}\label{sec:RSA}

\begin{figure}
\includegraphics[width=0.8\columnwidth]{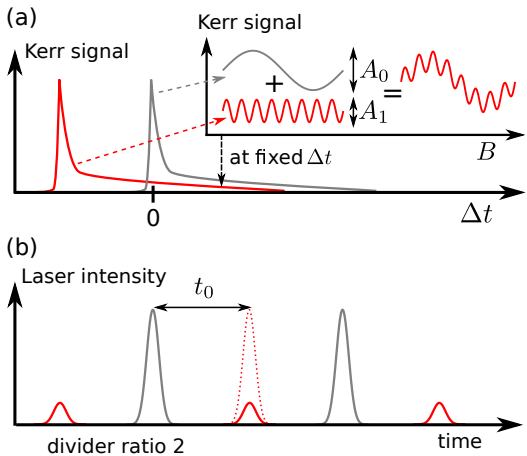}
\caption{\label{rsa-technique} (Color online) (a) A~schematic depiction of the resonant spin amplification technique (RSA). If the decay of spin population exceeds the time-spacing between neighbouring laser pulses, the MO signal measured for a~given value of a time delay $\Delta t$ is a sum of (at least) two signals. Inset: Dependence of MO signal for a~fixed value of $\Delta t$ on external magnetic field $B$. (b) A schematic illustration of the train of femtosecond laser pulses where each second laser pulse has a significantly lower intensity (in our case at least 30-times) due to the non-ideally working pulse picker which does not remove completely the laser pulse depicted by a dotted line (see Appendix~\ref{app:setup} for more details).}
\end{figure}

In our case, when the electron spin lifetime is longer than the time separation between adjacent pump laser pulses, the measured MO signal is a sum of signals coming from spin populations photoinjected by pump laser pulses at different times. The separation of these signals,
which is difficult to infer from a~time domain measurement, can be performed straightforwardly by the RSA technique\cite{kikkawa1998} that is schematically depicted in Fig.~\ref{rsa-technique}(a). The total Kerr signal $S(\Delta t,B)$ can be expressed as
\begin{eqnarray}
S(\Delta t, B) &= \sum_m \Theta\left(\Delta t +m t_0\right) A_m\mathrm{e}^{-\left(\Delta t +m t_0\right)/\tau_s}\nonumber \\ 
&\times\cos\left[g\mu_BB\left(\Delta t +m t_0\right)/\hbar\right],
\label{eq:RSA-all}
\end{eqnarray}
where $\Theta(x)$ is the Heaviside function that guarantees that only the previous pulses contribute to the total signal and $m=0,1,2\ldots$ is the index of the $m$-th preceding excitation pulse. $A_m$ is the initial amplitude of the Kerr signal at the instant of creation of the respective spin population by the $m$-th pulse (for equal intensities of all excitation laser pulses, $A_m=A$ is the same for all $m$). The term $g\mu_BB/\hbar$ is the Larmor frequency due to the external magnetic field $B$ applied in the sample plane where $g$, $\mu_B$ and $\hbar$ are the $g$-factor corresponding to the studied system, Bohr magneton and reduced Planck constant, respectively.

As follows from Eq.~\ref{eq:RSA-all}, if $\Delta t$ is fixed and $B$ is varied instead, two spin populations created by two subsequent excitation pulses contribute to the total Kerr signal with different oscillatory frequencies. This is due to a different time elapsed from the instant of the photoinjection of the corresponding spin populations [see Fig.~\ref{rsa-technique}(a)]. Consequently, the signal amplitudes corresponding to the different modulation frequencies are connected with the number of spins in these two populations. Therefore, one can probe simultaneously the time evolution of many mixed spin populations created at different instants by measuring the RSA curves at several $\Delta t$. 

The standard implementation of the RSA technique is based on an utilization of a~train of laser pulses with equal intensities.\cite{kikkawa1998} Here, on the contrary, we use a train of laser pulses where each second laser pulse has a significantly lower intensity using a non-ideally working pulse picker - see Fig.~\ref{rsa-technique}(b) and Appendix~\ref{app:setup}. The lower-intensity laser pulses create a~localized spin population at the GaAs/AlGaAs interface without disturbing the self-confinement, as discussed in Section~\ref{sec:separation}. A spin population created in such a way is already saturated with respect to the laser fluence, as can be seen in the inset in Fig.~\ref{rsa-ampl}(a).
This spin population serves as a probe of the influence of the much stronger (unattenuated) laser pulse which is incident on the sample at $\Delta t=0$~ps. We note that this modified RSA technique with unequal intensities of neighbouring laser pulses is, in certain sense, using the same idea as ordinary pump-probe experiment where a~weak probe pulse is used to measure the changes induced in the sample by a~strong pump pulse.

\section{Results}\label{sec:results}

\begin{figure}
\includegraphics[width=1.0\columnwidth]{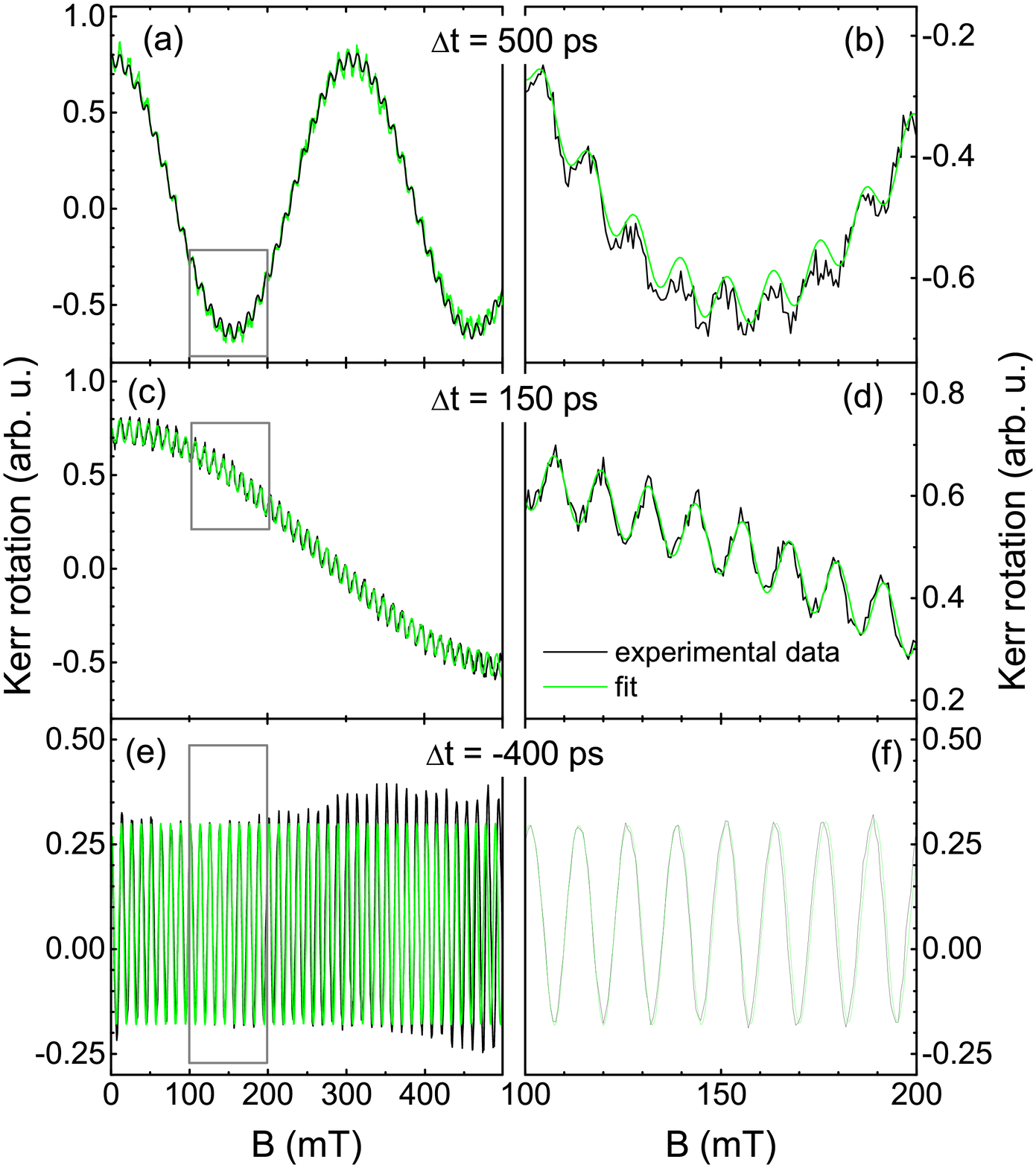}
\caption{\label{rsa-curves} (Color online) Results obtained by the RSA technique using a train of laser pulses where each second laser pulse has a significantly lower intensity. The experimentally measured data (black lines) were measured at $\Delta t=500$~ps (a,b), $150$~ps (c,d) and $-400$~ps (e,f); the right column show zoom views of the data marked by grey rectangles in the left column. The fits of the data by Eq.~\ref{eq:RSA-all} for $m=0$ and $1$ are plotted by the green (grey) curves. All data were measured at 10~K using laser fluence 140~$\mu$J$\cdot$cm$^{-2}$ and time separation of unattenuated pulses of $2 t_0=25$~ns.} 
\end{figure}

 In Fig.~\ref{rsa-curves} we show the results of our modified RSA technique for several values of time delays measured on the studied sample (see Sec.~\ref{sec:sample}). For $\Delta t=500$ and 150~ps, we can identify two signals with distinct frequencies - the larger signal oscillating with a smaller frequency corresponds to the spin population photoinjected by an unattenuated pulse at $\Delta t=0$ and the smaller signal oscillating with a larger frequency corresponds to the spin population photoinjected by a reduced-intensity pulse at $\Delta t= t_0=-12.5$~ns. On the other hand, for $\Delta t=-400$~ps [Fig.~\ref{rsa-curves}(e) and (f)] we see only one precession frequency because only the spin-population photoinjected by the reduced-intensity pulse is present in the sample for this time delay, which is in agreement with the deduced value of the spin lifetime $\tau_s\sim16$~ns$<2 t_0=25$~ns. Moreover, a strong reduction of the spin-population photoinjected by the reduced-intensity pulse due to the impact of the unattenuated pulse is immediately apparent from a comparison of the oscillation amplitudes in Fig.~\ref{rsa-curves}(f) and (d). To visualize this effect more clearly, we have fitted the measured data (see Fig.~\ref{rsa-curves}) by
Eq.~\ref{eq:RSA-all} for $m=0$ and 1.  

\begin{figure}
\includegraphics[width=1.0\columnwidth]{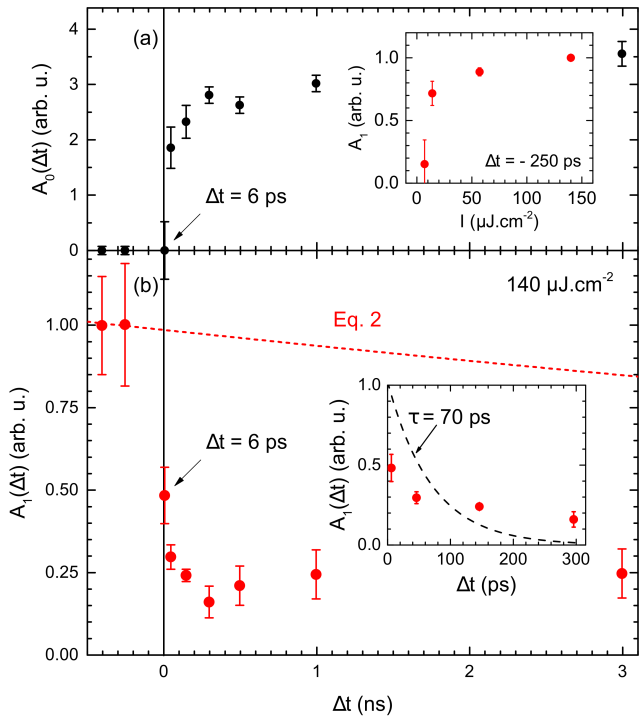}
\caption{\label{rsa-ampl} (Color online) Time evolution of spin populations photoinjected by laser pulses. (a) Amplitude $A_0(\Delta t)$ generated by the unattenuated pulse at $\Delta t=0$~ps. (b) Amplitude $A_1(\Delta t)$ generated by the diminished pulse at $\Delta t=-12.5$~ns. The red dotted curve depicts the expected exponential spin decay if no disturbing effects of the incident laser pulses is assumed (Eq.~\ref{eq:A1}). Insets: (a) The dependence of $A_1$ on the fluence measured at $\Delta t=-250$~ps. Due to a very weak (and not known precisely) intensity of the suppressed laser pulses, the x-scale is expressed in terms of the intensity of the unattenuated laser pulses, which is used to label the excitation conditions throughout the paper. (b) A zoom of data for small time delays with a calculated decay of spin population which would correspond to the electron-hole recombination with $\tau\approx70$~ps (black dashed curve). The experiment was performed at 10~K using laser fluence 140~$\mu$J$\cdot$cm$^{-2}$ and time separation of unattenuated pulses of $2 t_0=25$~ns.}
\end{figure}

If there were no disturbing effects of the incident laser pulses, the time delay-dependent amplitudes $A_m(\Delta t)$ should follow the exponential decay due to the electron spin relaxation
\begin{equation}
A_m(\Delta t)=A_m\mathrm{e}^{-\left(\Delta t + m t_0\right)/\tau_s}
\label{eq:A1}
\end{equation}
for $m=0$ and 1 according to Eq.~\ref{eq:RSA-all}. However, as illustrated in Fig.~\ref{rsa-ampl}(b), $A_1(\Delta t)$ does not follow the dependence expected from Eq.~\ref{eq:A1}, which is shown by the red dotted curve in Fig.~\ref{rsa-ampl}(b). Instead, the impact of the unattenuated pulse reduces it almost instantly, within $\Delta t<6$~ps, to one half of its value. Following this, the decrease of $A_1$ then slows down and it saturates at $\sim1/4$ of the original value within $\sim 1$~ns. Naturally, the amplitude $A_0(\Delta t)$ is zero for $\Delta t<0$ but, more interestingly, no spin population is injected to the sub-system within the first 6~ps after the laser pulse impact. In fact, it takes tens of picoseconds before the spin population starts to increase [Fig.~\ref{rsa-ampl}(a)] and it saturates slowly on a nanosecond time-scale.

We completed our time-resolved experiments with a time-averaged measurement of the nuclear spin polarization due to the presence of spin polarized photocarriers. The experiment is described in detail in App.~\ref{sec:nuclear-field}. By measuring the change of the Larmor frequency of electronic spins, we determined the effective magnetic field $B_n$ generated by spin polarized nuclei which adds to the external magnetic field. We used $B_n$ of completely saturated nuclei to estimate the time-average degree of spin polarization of electrons $P_e = (n_+ - n_-)/(n_+ + n_-)$ with respect to the average density $n_+$ or $n_-$ of electrons with their spin oriented parallel or antiparallel to the propagation vector of excitation light, respectively. We inferred $P_e\approx18$~\% and nuclei charging time $\approx60$ minutes.

\section{Discussion}\label{sec:discussion}
In the previous section we have shown experimental data demonstrating that a strong excitation pulse triggers a very fast reduction of the existing spin population $A_1$. In the following text we discuss possible mechanisms that can be responsible for this observation. 

 

First, we describe the processes which take place if a train of circularly polarized laser pulses is absorbed in an $n$-type doped bulk semiconductor. Due to the optical orientation, each laser pulse excites a non-equilibrium population of spin-polarized electrons and holes which recombine with a time constant $\tau$. Thanks to the presence of equilibrium electrons in the conduction band, which are provided by the doping, a certain fraction of the photo-injected holes recombines with these pre-existing electrons. Consequently, the remaining photo-injected spin-polarized electrons can maintain their spin for time $\tau_s$ significantly exceeding $\tau$.\cite{dzhioev2002,kikkawa1998} For $\tau_s$ larger than the spacing between the adjacent laser pulses $t_0$, the resulting electronic spin population is always partially reduced by an impact of the following laser pulse due to a recombination with photo-injected holes and new spin-polarized electrons are added, which is leading to a periodic spin renewal. 


In our sample, the observed fast reduction of the spin population $A_1$ [to $\approx50\%$ of the initial value in time $<6$~ps as shown in the inset in Fig.~\ref{rsa-ampl}(b)] cannot be explained by the recombination of the existing electrons with the photo-injected holes. This is because we observed experimentally that $\tau=70$~ps  [see the inset in Fig.~\ref{sample-Kerr}(c) for the transient reflectivity decay measurement], a value consistent with the literature\cite{kikkawa1998, wu2010,nemec2005}. Moreover, a recombination of electrons would free some states in the studied self-confined sub-system, which are nearly fully occupied by electrons for our experimental conditions [see the saturation behavior of the MO signal with the pump intensity in the inset in Fig.~\ref{rsa-ampl}(a)]. Consequently, the fast reduction of $A_1$ due to a recombination of electrons should be accompanied by an increase of newly injected population $A_0$ into the sub-system, which is not the case (see Fig.~\ref{rsa-ampl}(a) where $A_0\approx0$ for $\Delta t=6$~ps).

Another possible mechanism, which could in principle reduce $A_1$ without a decay of the electron concentration, is the spin relaxation process. The most efficient mechanism in GaAs at low temperatures, especially in confined systems, is the Dyakonov-Perel (DP) spin dephasing.\cite{zutic2004} Indeed, the DP mechanism can be the origin of a fast spin relaxation with $\tau_s$ of tens or even units of ps in systems with a strong confinement such as remote-doped AlGaAs/GaAs heterointefaces or quantum wells (see, e. g., Ref.~\onlinecite{wu2010} and references herein). However, we observe exceptionally long $\tau_s\approx16$~ns which is a signature of a weak confinement and an inefficient DP dephasing in the studied sample.\cite{nadvornik2016} The photoexcitation of carriers due to the absorption of an intense excitation pulse can be expected to weaken the confinement [see Fig. 1(b)]. Moreover, the eventual pump-induced fluctuations of the confinement would lead to a further decrease of the DP efficiency due to the motional narrowing \cite{zutic2004,nahalkova2006} and, thus, it would lead to a further increase of $\tau_s$. Overall, this rules out the DP and other less efficient spin-relaxation mechanisms as a cause of the fast decay of $A_1$.


We attribute the observed phenomenon to a removal of the spin-polarized electrons from the self-confined electron sub-system. This is due to the pump-induced disturbance of the confining potential and, thus, to a reduction of the density of available states in the sub-system. The confinement is a self-consistent process which depends strongly on boundary conditions and local charge distribution and can be disturbed by a high density of photo-injected charges.\cite{nadvornik2016} When the confining potential is modified by the additional carriers in the bulk GaAs, the excess electrons move from the sub-system to the bulk [see Fig. 1(b)]. Therefore, they do not contribute significantly to the measured MO signals, which were optimized for the sub-system detection as discussed in Sec.~\ref{sec:typical-kerr} and in more details in App.~\ref{app:setup}. 
In the following tens and hundreds of picoseconds, the electrostatic confinement starts to recover due to a continuous decay of photo-carriers in the bulk via their recombination and, consequently, additional states become available in the sub-system. These new states and the states made available by a partial recombination of $A_1$ electrons with the photo-injected holes are subsequently filled by spin-polarized electrons from the bulk. This results in an increase of the spin-polarized population $A_0$ within $\approx300$~ps (see Fig.~\ref{rsa-ampl}). To sum up, this experiment revealed that the spin polarization can be periodically ``erased'' from the superior highly mobile spin-transport layer in times $<6$~ps and regenerated again in hundreds of picoseconds. This defines the constraints for the high-rate operation of spin-logic devices based on this structure.


A side effect of this functionality is an unideal renewal of the spin system due to the incomplete decay of $A_1$ and other possible perturbations of the spin system. This leads to a decrease of the degree of spin polarization $P_e$ which is practically achievable in the sub-system. 
Ideally, each absorbed laser pulse should generate $P_{e}\approx50$\% at the instant of absorption\cite{bhat2005}. If a material with $\tau_s\approx16$~ns is excited by a train of circularly polarized laser pulses with a repetition rate of 80~MHz it should lead to time-average $P_e\approx36$\% when this unideal renewal is not considered. Experimentally, we obtained $P_e\approx18$\% that shows, indeed, a visible but not dramatic reduction of the spin polarization.

\section{Conclusion}
Long-range and high-speed spin transport together with fast spin manipulation are the keystones of high-operation-rate spin-logic devices. In this paper we show that the fast spin removal functionality can be added to a superior spin-transport channel in undoped GaAs/\algaas\ heterointerfaces using absorption of optical pulses. By employing a modified resonant spin amplification technique with unequal intensities of neighbouring laser pulses we demonstrated that spins can be removed from the high-performance self-confined system in a picosecond timescale by optical manipulation of the densities of carriers in the system. This is achieved by reducing the degree of confinement due to the photo-injection of carriers in this region. It was also observed that the recovery of the confinement occurs in a few hundreds of picoseconds. The demonstrated functionality which allows for a fast periodic erasing and regenerating of the electronic spin polarization in a long-range and high-speed spin-transport layer makes the system an excellent candidate for fast spin-logic devices. Since these superior properties should be, in principle, common for a larger variety of self-confined semiconductor structures, the present finding can open new perspectives for spin-logic applications.


\begin{acknowledgments}
We acknowledge support from the European Research Council (ERC) Advanced Grant No. 268066, from European Metrology Research Programme within the Joint Research Project EXL04 (SpinCal), from the Ministry
of Education of the Czech Republic Grant No. LM2015087, from the Czech Science Foundation Grant No. 14-37427G, from the Charles University Grants No. 1582417 and No. SVV-2015-260216.
\end{acknowledgments}

\appendix
\section{Experimental setup}\label{app:setup}
The used time-resolved spin-injection and detection method relies on the pump-probe (P\&P) technique\cite{kimel2000}. The spin-injection is realized using the optical orientation\cite{agranovich1984} by the circularly polarized excitation laser pulse and the spin-detection is done by measuring the polarization rotation of the linearly polarized probe pulse via the magneto-optical polar Kerr effect\cite{korn2010}. 

\begin{figure}
\includegraphics[width=0.8\columnwidth]{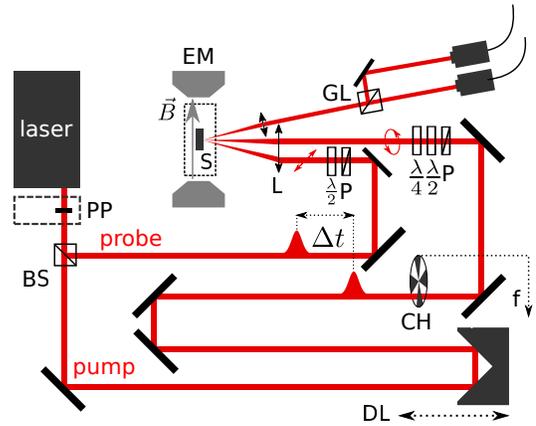}
\caption{\label{setup} (Color online) A~Schematic depiction of the used P\&P magneto-optical setup. Femtosecond laser pulses were splitted by a~beam splitter (BS) to pump and probe pulses with a~delay line (DL) controlled mutual time delay ($\Delta t$). The required polarization state of laser pulses was set using a~combination of polarizers (P) and waveplates ($\lambda/2$ and $\lambda/4$) and they were focused by a~single lens (L) on the sample (S) placed in a~cryostat. The reflected probe pulses were collimated by another lens and their polarization rotation was measured by an optical bridge based on a~Glan-Laser (GL) polarizer an a~pair silicon photo-diode detectors. The difference electrical signal was processed by a~lock-in amplifier at a~modulation frequency $f$ of the optical chopper (CH). External magnetic field $B$ was applied in the sample plane by an electromagnet (EM). Optionally, the pulse-picker (PP) can be used to increase a~time separation between neighbouring laser pulses.}
\end{figure}

A schematics of the experimental setup is shown in Fig.~\ref{setup}. A~mode-locked Ti:sapphire laser (Mai Tai, Spectra Physics) with a~repetition rate 80~MHz (i.e. a~time separation between neighbouring laser pulses $ t_0=12.5$~ns) was set to wavelength $\lambda=815$~nm, close to the bandgap of GaAs at low temperatures.\cite{surynek2017} The generated $\sim100$~fs  laser pulses were splitted to pump and probe pulses and their fluence was adjusted by neutral density filters (their intensity ratio at the sample was typically 10:1). A time delay between pump and probe pulses $\Delta t$ was controlled by a~delay line from -0.5~ns to 3.5~ns. The laser pulses were focused by a~converging 10D lens on the sample to $\sim25$~\mum-sized overlapped spots. Unless explicitly mentioned otherwise, the angles of incidence to the sample surface were $<1^{\circ}$ and $\approx7^{\circ}$ for the pump and probe beams, respectively. The sample was mounted in an optical cryostat at temperature of $~10$~K and placed between the poles of an electromagnet that provided in-plane magnetic field up to $B=500$~mT oriented in the plane of light incidence. The MO signal corresponding to the probe polarization rotation and the differential reflectivity were measured as a difference and sum signal in the optical bridge, respectively.\cite{rozkotova2008}

A pulse picker (PP) can be inserted to the setup at the position marked by ``PP'' in Fig.~\ref{setup}. A PP is essentially an electrically controlled acousto-optical modulator that transmit only certain pulses and block all the others. The PP in MO experiments usually used in order to determine precisely the spin lifetime from MO time-resolved signals in systems where the spin polarization decay time exceeds the time separation $t_0$ between the neighbouring laser pulses. The dilution of laser pulses achieved by the PP is characterized by a so-called divider ratio $n$ which describes that the time separation between neighbouring pulses is increased to $n t_0$. However, in the case of our experiments, we employed the PP in order to modulate the intensity of pulse train instead of the selection of every $n$-th pulse. This is possible thanks to the unideal efficiency of the pulse suppression by the PP. We verified experimentally, that the diminished laser pulses has intensity suppressed at lest 30-times, which is the detection limit given by the noise level in our measurement with a fast photodiode, with respect to that of the unattenuated ones. The diminished laser pulses create the spin-polarized population in the sub-system which is already saturated with respect to the laser fluence [see the inset in Fig.~\ref{rsa-ampl}(a)]. This means that all states which are available in the confined system are already filled by the diminished excitation pulse.


The polar Kerr efect, which is the origin of the detected probe polarization rotation in our sample, changes a sign around the semiconductor bandgap - see inset in Fig. 1 in Ref.~\onlinecite{crooker2005}. Consequently, if spectrally broad laser pulses are used, which is the case for our $\sim100$~fs long laser pulses (with a spectral bandwidth of $\sim10$~nm), the measured MO signals can be positive, negative or even close to zero depending of the mutual spectral position of the laser pulse and the MO spectrum. If more than one MO-active system with a distinct MO spectrum is present in the sample, the fine tuning of the probe pulses can be used to enhance or suppress the sensitivity of the detected MO signal to the system of interest. This is the reason why the spins photoinjected to the bulk region outside the confined sub-system, which are rather short-lived, do not significantly contribute to the MO signals shown in this paper, which were optimized with respect to the confined electron sub-system - see Fig.~\ref{sample-Kerr}(c) and Fig.~\ref{rsa-ampl}(a). More discussion about this can be found in the Supplementary information in Ref.~\onlinecite{nadvornik2016}.

\section{Estimate of degree of electron spin-polarization from nuclear polarization measurements}\label{sec:nuclear-field}
In semiconductors whose nuclei carry a magnetic moment, like GaAs, the degree of spin-polarization of electrons, $P_e$, can be transferred from electrons to nuclei via the hyperfine interaction. The evolution of the degree of polarization of nuclei $P_n$ in the laboratory time $t$ follows the exponential function\cite{greilich2007,shiogai2012,hashimoto2002}
\begin{equation}
P_n(t)\propto P_e\left(1-\text{e}^{-t/T_n}\right),
\label{eq:charging}
\end{equation} 
where $T_n$ is the characteristic time of the electron-nuclear hyperfine coupling, which is usually of the order of minutes\cite{greilich2007,shiogai2012,hashimoto2002,paget1982} (i.e., much longer that the electron spin relaxation time). Consistently with the main text, we use the standard definition of the degree of polarization,\cite{zutic2004} $P = (n_+ - n_-)/(n_+ + n_-)$, with respect to the density $n_+$ or $n_-$ of spin-carriers with a~spin oriented parallel or antiparallel to the quantization axis (propagation vector of the circularly polarized excitation beam). 

\begin{figure}
\includegraphics[width=1.0\columnwidth]{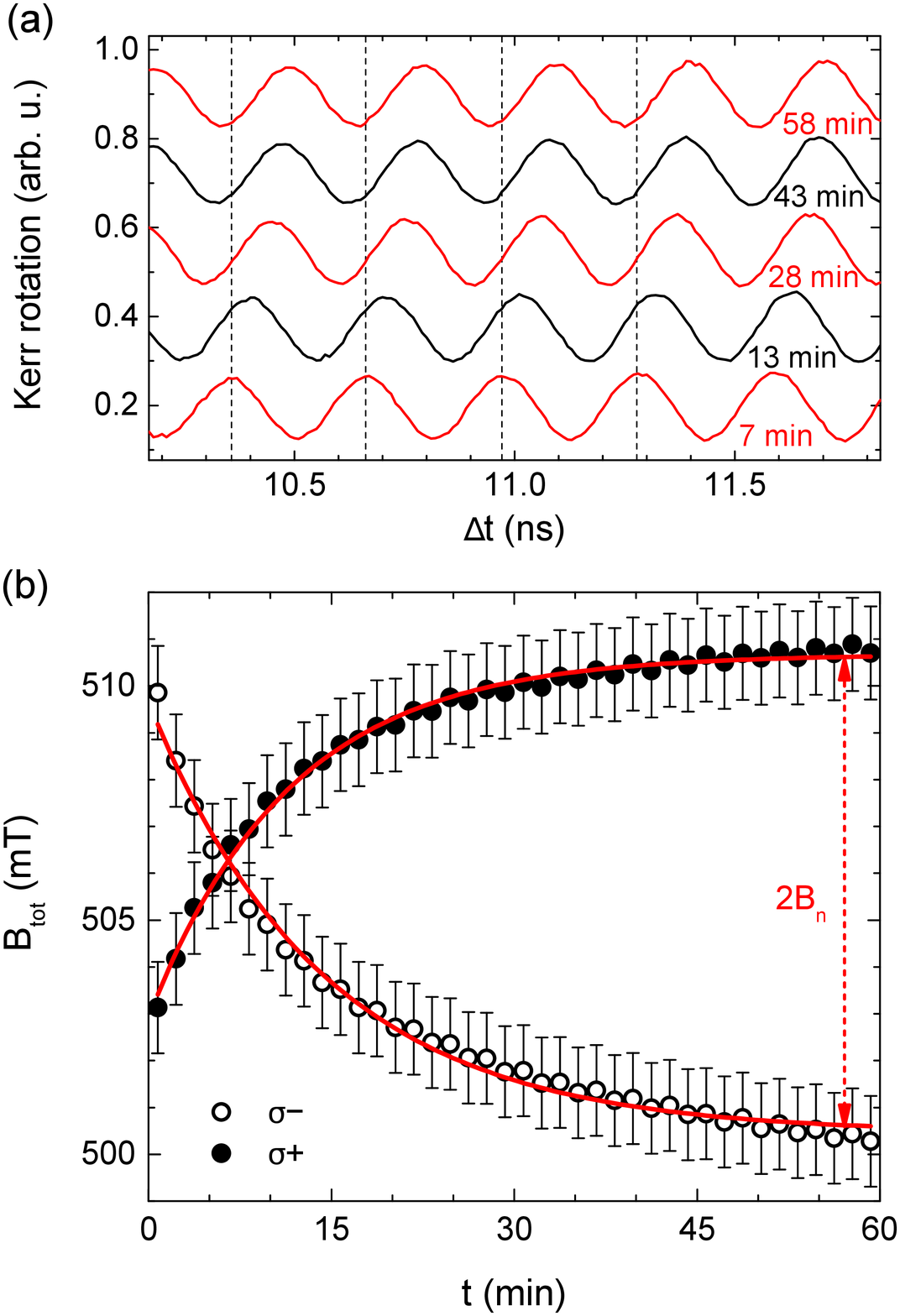}
\caption{\label{nuclear} (Color online) (a) Time-resolved MO signal measured on the sample whose nuclei have been spin-polarized by a long term exposure to light with the helicity $\sigma_+$. The curves and the corresponding labels indicate the laboratory time during which the sample was illuminated by light with the helicity $\sigma_-$; the curves were vertically shifted for clarity. The vertical lines emphasize the variation of the precession frequency due to the varying nuclear magnetic field. Data were measured for the time separation between neighbouring pulses $n t_0=12.5$~ns, excitation fluence 14~$\mu$J$\cdot$cm$^{-2}$, sample temperature $T=10$~K, and external magnetic field $B=500$~mT. (b)~The dependence of $B_{\text{tot}}=B+B_n$ on laboratory time for excitation by helicities $\sigma_-$ (open symbols) and $\sigma_+$ (closed symbols) measured on the sample whose nuclei have been spin-polarized by a long term exposure to light with the opposite helicity. The curves are the fits by Eq.~\ref{eq:charging2}. The vertical arrow depicts the saturated value of $2B_n$.}
\end{figure}

As the nuclei are getting polarized, they generate a nuclear magnetic field, $\vek{B}_n=b_n\left\langle \vek{P}_n\right\rangle$, that adds to the external magnetic field $\vek{B}$. Here, $b_n=-8.5$~T is the magnetic field that would be generated by fully spin-polarized nuclei and $\vek{P}_n=P_n\vek{\hat{I}}$ is the vector of nuclear spin-polarization oriented along the unit vector of the nuclear spin $\vek{\hat{I}}$. When $P_n$ achieves its saturation value for $t\gg T_n$, $\vek{B}_n$ can be expressed using $P_e$ as\cite{paget1977,salis2009} 
\begin{equation}
\vek{B}_n = fb_n^{\ast}\frac{\vek{B}\cdot\vek{P}_e}{B^2}\vek{B},
\label{eq:Bn}
\end{equation} 
where the vector $\vek{P}_e=P_e\vek{\hat{s}}$ is oriented along the unit vector of the electronic spin $\vek{\hat{s}}$ ($P_e=1$ for fully spin-polarized electronic system), $b_n^{\ast}=-8.5$~T is the effective magnetic field generated by a~spin-polarized nuclei\footnote{The value of $b_n^{\ast} = -8.5$~T is valid for $P_e\lesssim 0.5$ and it decreases to $-5.4$~ T for fully spin-polarized nuclei (i. e., for $P_e = 1$). This comes from the fact that a~spin of a~nucleus is a~5 level system (while electronic spin is a~2 level system) and its total magnetic moment is not a~linear function of $P_e$.\cite{dyakonov1974}}
 and $f\leq 1$ is a~phenomenological leakage factor\cite{paget1977} that takes into account possible processes of nuclear spin-relaxation other than the hyperfine coupling with electrons. Usually, it is reported $fb_n^{\ast}\approx -0.85$~T.\cite{paget1977,shiogai2012,christie2015,sallen2014} Eq.~\ref{eq:Bn} holds for $B\gg1$~mT when the nuclear dipole-dipole interactions are negligible. Consequently, if the angle between $\vek{B}$ and $\vek{P}_e$ is known, the value of $P_e$ can be inferred from the measured value of $B_n$. 

In our experimental setup, the angles of incidence to the sample surface were $<1^{\circ}$ and $\approx7^{\circ}$ for the pump and probe beams, respectively (see Appendix~\ref{app:setup}). However, the polarization of nuclei happens only if there is a non-zero component of the optically injected spins in the direction of the magnetic field, lying in the sample plane in our experimental configuration (see Eq.~\ref{eq:Bn}). So, the angle between $\vek{B}$ (the sample plane) and $\vek{P}_e$ (the direction of the pump beam) has to be different from $90^{\circ}$ - i.e., the angle of incidence of the pump beam has to be non-zero for an efficient polarization of the nuclei. To achieve this, we exchanged the role which the optical beams play in our setup (see Fig.~\ref{setup}). Consequently, the angle of incidence of the pump beam in this configuration was $\approx7^{\circ}$ and the range of accessible time delays was shifted to $\Delta t = 9-13$~ns.  

Since the spin transfer from electrons to nuclei via the hyperfine channel is a rather slow process whose characteristic time $T_n$ is of the order of minutes or tens of minutes,\cite{greilich2007,shiogai2012,hashimoto2002,paget1982} the process of nuclei polarization can be measured directly in the laboratory time. As an example, in Fig.~\ref{nuclear}(a) shows a sequence of time-resolved MO signals measured for $\sigma_-$ pump helicity at depicted times after a long term sample exposure to light with the helicity $\sigma_+$. Clearly, we observe a gradual change of the precession frequency $\Omega=g_s\mu_BB_{\text{tot}}/\hbar$ due to the increase of the total magnetic field $B_{\text{tot}}=B+B_n$ experienced by spin-polarized electrons. In Fig.~\ref{nuclear}(b) we show the values of $B_{\text{tot}}$ inferred from the fits of measured dynamics as a function of the laboratory time for both helicities of pump pulses. We observe that the nuclear polarization saturation occurs after $t\approx 60$~minutes. This time is roughly one order of magnitude longer than the values reported in literature for $n$-doped GaAs (with $n\approx 2-5\times 10^{16}$~\cmThree).\cite{shiogai2012,paget1982} Considering that charging times longer than tens of minutes are reported in undoped GaAs,\cite{hashimoto2002} these data suggest a low density of the spin-polarized electron sub-system in agreement with the concentration of the order of $\sim10^{15}$~\cmThree\ which was determined experimentally for this sample in Ref.~\onlinecite{nadvornik2016}.

Following Eq.~\ref{eq:charging}, the precise value of $B_n$ of the saturated nuclear polarization, which is depicted as a vertical arrow in Fig.~\ref{nuclear}(b), can be determined by fitting the deduced values of $B_{\text{tot}}$ for both helicities by 
\begin{equation}
B_{\text{tot}}(t) = B_{\text{tot}}^{\pm} \pm C^{\pm}\mathrm{e}^{-t/T^{\pm}_n}. 
\label{eq:charging2}
\end{equation}
Here, $B_{\text{tot}}^{\pm}$ are the saturation total magnetic fields for the excitation with $\sigma_{\pm}$, $C^{\pm}$ are the corresponding amplitudes of the charging process, and $T^{\pm}_n$ are the characteristic times. We note that $\vek{B}_n$ is parallel or antiparallel to $\vek{B}$ with respect to the excitation by $\sigma_+$ and $\sigma_-$. Considering the error bars, the fits give $B_n=(B_{\text{tot}}^+-B_{\text{tot}}^-)/2=5.1\pm0.5$~mT and the average $T_n=(13\pm1)$ minutes. Finally, the degree of electron spin-polarization $P_e\approx18\%$ is determined using Eq.~\ref{eq:Bn} and by considering the refraction of the pump light inside the GaAs sample due to its the refractive index of $n=3.67$ at $\lambda = 815$~nm.

%

\bibliography{Refs_LN}%


\end{document}